\documentclass{eas}
\usepackage{graphicx}
\pdfoutput=1
\begin{document}

\title{Status and Results from AMANDA/IceCube} 
\runningtitle{AMANDA/IceCube}
\author{Patrick Berghaus for the IceCube Collaboration}\address{For a full collaboration list: http://www.icecube.wisc.edu}
\begin{abstract}
IceCube is a cubic kilometer-scale neutrino telescope under construction at the South
Pole since the austral summer 2004/2005.
At the moment it is taking data with 22 deployed strings. The full
detector is expected to be completed in 2011 with up to 80 strings each
holding 60 digital optical modules.
The progenitor detector AMANDA has been operating at the same site
since 1997 and is still running as an integral part of IceCube.
A summary of AMANDA science for its 10 years of standalone operations is
presented, as well as the status and first physics results of the
IceCube project.
\end{abstract}
\maketitle
\section{Introduction}
\subsection{High-Energy Neutrino Astrophysics}

Neglecting gravity, neutrinos interact only by way of the weak interaction. This precludes absorption by matter and radiation fields which affect photons and therefore potentially allows observation of objects and processes that are not accessible to conventional gamma-ray astronomy. Targets of interest include the neighborhood of black holes and active galaxies, photons from where are subject to absorption by ambient matter and extragalactic background light. The obvious disadvantage resulting from the restriction to weak interactions is the need for very large detector volumes to achieve at least a moderate event yield. 

One of the longest-standing problems in astrophysics is the search for the origin of cosmic radiation \cite{Halzen:2002pg}. Since the direction of charged cosmic rays, except at the very highest energies, effectively gets scrambled by magnetic fields, any information about their origin must come from electrically neutral particles, i.e. photons and neutrinos. Investigation of this issue is one of the main goals of Very High Energy (VHE) neutrino detectors, with instrumented volumes upwards of $10^{6}{\rm m}^{3}$. These take advantage of Cherenkov radiation of charged particles produced in interactions of neutrinos with ambient matter. The large fiducial volume required in the detection of the comparatively low neutrino fluxes at TeV-PeV energies makes it necessary to take advantage of large natural reservoirs of light-transparent media, such as the deep sea or polar ice sheets.

So far, no VHE neutrino signal from outside the solar system has been detected. This situation is expected to change after completion of a new generation of detectors with an instrumented volume of the order of $1{\rm km}^3$ \cite{Halzen:2007ip}.

\begin{figure}
\includegraphics[width=3in]{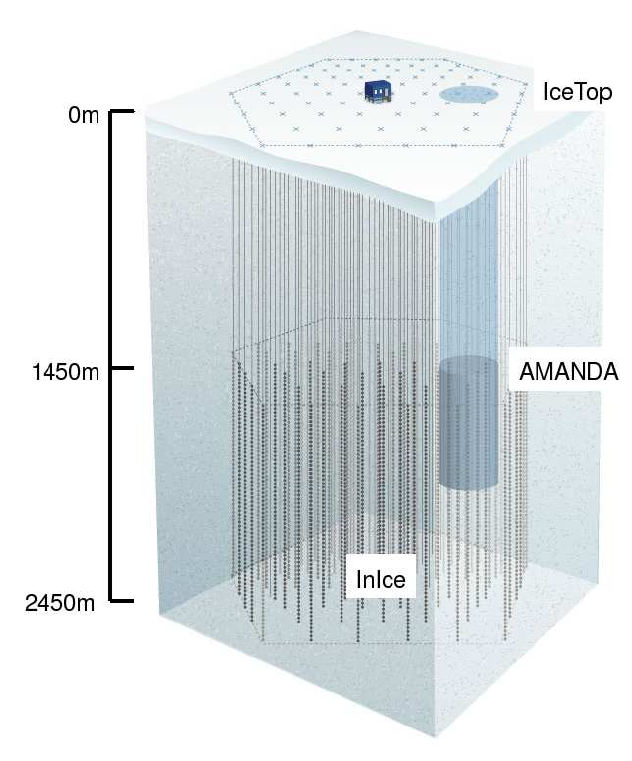}
\caption{The three components of IceCube: IceTop (surface array), InIce and AMANDA.}
\end{figure}

\subsection{The IceCube Neutrino Observatory}

IceCube is expected to become the first operational cubic-kilometer scale neutrino detector. Built at the geographic South Pole around its predecessor AMANDA, the sub-surface detector (InIce) is scheduled to reach its full extent of 80 instrumented strings carrying 60 digital optical modules (DOMs) each by the year 2011 \cite{Hill:2006mk}. Already, with 22 deployed strings, it is the largest neutrino telescope ever built and has started delivering data of unprecedented quality. An additional detector component, the surface air-shower array IceTop, will eventually comprise 80 pairs of frozen water tanks with two DOMs in each.

Physics topics in IceCube, other than astrophysics, include investigation of neutrino oscillations, indirect searches for dark matter and exotic particle physics, such as the search for magnetic monopoles. Due to its location, IceCube will be sensitive mostly to neutrino signals from the northern hemisphere, since above the horizon background from downgoing muons makes detection of neutrino events very challenging.

\subsection{Phenomenology of $\gamma$-Ray Production}

Processes leading to the production of VHE $\gamma$-radiation can be subdivided into two classes. In \emph{electronic} mechanisms, gamma rays are produced by inverse Compton scattering of high energy electrons on ambient photons. Conversely, \emph{hadronic} models assume gamma production through the decay of neutral pions, which in turn are produced in interactions involving high-energy protons. Observations by gamma-ray telescopes have so far not provided unambiguous evidence for either model, irrespective of the source type \cite{Aharonian:2007bn}.

Neutrino astrophysics provides a natural means to distinguish between the two mechanisms, since hadronic interactions will also produce charged mesons, all of which decay into final states containing at least one neutrino. Purely electronic emission processes produce neutrinos only in negligible quantities. Any detection of neutrinos from a specific source would therefore provide very strong evidence for the hadronic model \cite{Halzen:2007ip} and consequently allow to identify the sources of cosmic radiation.

\section{AMANDA Results}

\begin{figure}
\includegraphics[width=5in]{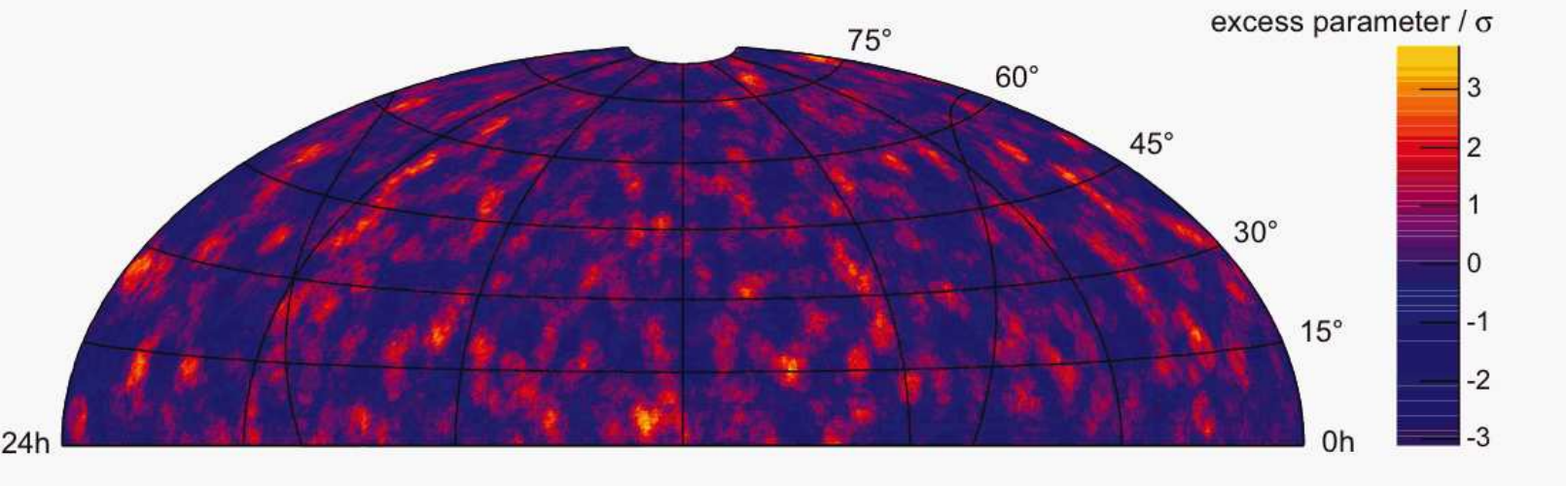}
\caption{Significance map for 5-year AMANDA-II point-source search. The maximum value of $3.7\sigma$ excluding trial factors is fully consistent with random statistical fluctuations.}
\end{figure}

\subsection{Astrophysical Point Sources}
The search for point sources of extraterrestrial neutrinos has been a major analysis topic in AMANDA. For this purpose, muon neutrino candidate events originating from the northern hemisphere have been used. As of the time of writing, data from all years of AMANDA operation have been analyzed in separate searches. The best published limit so far covers the years 2000-2004. Using 4282 candidate events, the limit for the muon neutrino flux from point sources averaged over the northern hemisphere is $E_{\nu}^{2}d\Phi_{\nu_{\mu}}/dE=5.5\times10^{-8}GeVcm^{-2}s^{-1}$ in the energy range between 1.6TeV and 2.5PeV \cite{Achterberg:2006vc}. Here, as in all subsequent results, a spectral index of $\gamma=-2$ was assumed for the signal.

Recently, for the first time a combined point-source limit was calculated for the years 1997-99, during which AMANDA was operating effectively in a 10-string configuration. The result from this analysis, using 465 candidate events, is $E_{\nu}^{2}d\Phi_{\nu_{\mu}}/dE<5\times10^{-7}GeVcm^{-2}s^{-1}$. Also lately, data from the 2005 observation period were analyzed using a likelihood-based method \cite{jim_braun}, yielding an improvement of 30\% in both sensitivity and discovery potential over the old technique. None of the samples obtained so far shows any statistically significant excess over background expectation. A combined result using all available AMANDA-II data will be published soon.

\subsection{Diffuse Neutrino Flux}
The diffuse analysis represents an alternative approach in the search for a cosmic neutrino signal. Here, an all-sky search is conducted looking for an excess at high neutrino energies. To separate the signal from background, an energy-correlated parameter cut is applied after carefully cleaning the data of downgoing muon tracks. For AMANDA data from 2000 to 2003, this method yielded 6 events in the final sample, compared to an expected background of 6.1. This corresponds to a limit for diffuse muon neutrino flux from the northern hemisphere of $E^2\Phi<8.8\times10^{-8}GeVcm^{-2}s^{-1}sr^{-1}$ for an assumed $E^{-2}$ spectrum with $15.8{\rm TeV}<E_\nu<2.5{\rm PeV}$, covering 90\% of the simulated signal \cite{diffuse}. The result also has implications for prompt (charm) production of atmospheric neutrinos, ruling out some of the highest-yielding models at 90\% C.L..

\subsection{Other AMANDA Results}
Searches for neutralino WIMP signals from the Sun and Earth have been completed using data from 2001 and 2001-2003 respectively \cite{Achterberg:2006jf}. The limits on the muon flux from neutralino annihilation are consistent with or slightly better than those from other indirect searches, in spite of AMANDA's shorter integrated live time. Analyses for the remainder of AMANDA data are ongoing \cite{daan}.

Gamma Ray Bursts (GRB) are believed to be a likely source for ultra-high energy cosmic rays and high energy neutrinos, motivating a search for a neutrino signal in coincidence with transient gamma signals from satellite-borne detectors. For the years 2000-2003, 407 bursts were analyzed looking for corresponding up-going muon tracks, but yielding no neutrino candidate events inside the temporal and spatial search windows. Also, ``rolling'' searches were conducted looking for an excess of cascade events within a predefined time independent of known burst alerts. All results were consistent with background expectation. The best limit so far, obtained by the coindicent muon analysis, lies slightly above the Waxman-Bahcall flux, one of the standard benchmarks in calculating neutrino yields from GRB \cite{grbpaper}.

Further studies have been undertaken looking for Ultra-High Energy (UHE) neutrinos \cite{lisa_gerhardt}, magnetic monopoles \cite{hickes} and diffuse neutrino fluxes based on detection of electromagnetic and hadronic cascades from $\nu_e$, $\nu_{\tau}$ and neutral current interactions\cite{oksana}. None of the results showed a statistically significant excess over background.

\section{IceCube}
\subsection{IceCube Status}
After the deployment period during austral summer 2006/07, a total of 22 strings have been completed containing 1424 DOMs, 97.6\% of which are fully functional as of summer 2007. The trigger rate in this configuration is 550Hz, and expected to rise to 1650Hz in the full 80-string detector\cite{all_bracked}. In addition to the InIce strings, 26 IceTop surface tanks are operational, each containg two DOMs. Construction of the detector is proceeding according to schedule and is expected to be completed by 2011.

\begin{figure}
\includegraphics[width=4in]{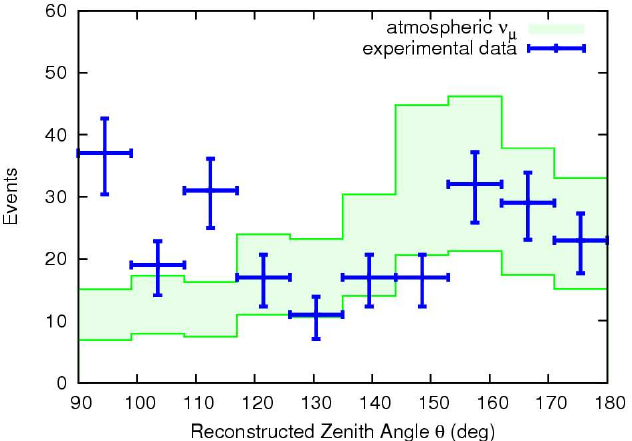}
\caption{Zenith angle distribution of IceCube-9 atmospheric neutrino events after final cuts compared to Monte Carlo simulation.}
\end{figure}

\subsection{First IceCube Results}
Using data from the first nine strings during the 2006 observation period, several analyses have been conducted, largely with the goal of validating detector performance. An important result was obtained in the search for atmospheric neutrinos. A total of 234 candidate events were identified, compared with an expectation of $211\pm76.1(syst.)\pm14.5(stat.)$. While there is some indication of residual contamination by downgoing atmospheric muons near the horizon, at about 30 degrees below the distribution becomes fully consistent with a pure neutrino sample \cite{Achterberg:2007bi}.

A point source search was conducted using the same data set. The sky-averaged sensitivity for an $E^{-2}$ spectrum, $E^{2}_{\nu}d\Phi_{\nu_{\mu}}/dE=1.2\times10^{-7}GeVcm^{-2}s^{-1}$, is comparable to one year of AMANDA-II operation. Even with only nine strings, IceCube already has a higher effective area at high neutrino energies than AMANDA-II\cite{chad}. Results from searches for diffuse neutrino signals are expected to be published soon \cite{aya}.

\subsection{Outlook}
As the IceCube detector increases in volume, new possibilities for physics analyses will arise. One entirely new path of investigation is the search for tau neutrino interactions. Due to flavor oscillations, one third of all astrophysical neutrinos arriving at Earth is expected to be a $\nu_{\tau}$. For this type of interaction, several characteristic signatures have been identified. The major advantage in detecting tau neutrinos is the very low background from atmospheric interactions, making detection of an extraterrestrial signal significantly less ambiguous that in other channels \cite{Cowen:2007ny}.

Operation of the AMANDA detector as a subsystem of IceCube is expected to continue for the forseeable future. As a consequence, neighboring IceCube strings can be used to veto tracks coming from above the horizon, extending the active aperture for detection of muon neutrinos to the southern hemisphere. Since this is where the bulk of the galactic plane, and hence the majority of potential galactic neutrino sources, is located, it can be expected to have a significant impact on science operations. Furthermore, with background vetoing, the effective energy threshold should decrease to values below 50GeV \cite{big_andy}.

Several analyses will make use of the entire combined IceCube system. The combination of AMANDA and InIce will allow to set more stringent limits on solar WIMPs. Preliminary investigations indicate an improvement by up to an order of magnitude for low mass neutralinos, in part resulting from the recent introduction of a new Transient Waveform Recording (TWR) DAQ system in AMANDA \cite{gustav}. The IceTop surface array together with the InIce component should be able to study cosmic rays up to energies of 1EeV. Combination of surface and depth measurements of muon showers will allow investigation of cosmic ray composition, especially around the ``knee'' at $E_{primary}=3{\rm PeV}$ \cite{icetop}.

Various techniques for multi-messenger campaigns have been proposed, among them the use of ``Maximum Likelihood Blocks'', preliminarily demonstrated on blazar X-ray data \cite{Resconi:2007nx}. Also, a test run for Target of Opportunity observations in coordination with the MAGIC VHE-gamma detector was recently concluded \cite{markus}, and the beginning of a regular campaign is expected for the near future.

\section{Conclusion}

AMANDA has during its time of operation become the world's largest and most sensitive astrophysical neutrino detector at VHE energies and above. Results for a wide variety of physics investigations have been published, with several more in preparation.

The transition to IceCube has so far been remarkably successful. First physics results have been published, validating the excellent performance of the new detector. As IceCube continues to grow with every deployment season, increases in both volume and angular resolution should soon allow results to significantly surpass those from AMANDA. An important symbolic milestone for IceCube will be reached in 2008, as integrated exposure will exceed $1{\rm km}^3{\rm yr}$. Data becoming available over the course of the next few years will allow important physics investigations, such as a probe of the diffuse Waxman-Bahcall flux and exclusion or confirmation of various emission models for individual cosmic objects.



\begin{thebibliography}{99}

\bibitem[(Achterberg {\it et al.} 2006)]{Achterberg:2006jf}
  A.~Achterberg {\it et al.}  [AMANDA Collaboration],
  Astropart.\ Phys.\  {\bf 26} (2006) 129.

\bibitem[(Achterberg {\it et al.} 2007a)]{Achterberg:2006vc}
  A.~Achterberg {\it et al.}  [IceCube Collaboration],
  Phys.\ Rev.\  D {\bf 75} (2007) 102001
  [arXiv:astro-ph/0611063].

\bibitem[(Achterberg {\it et al.} 2007b)]{Achterberg:2007bi}
  A.~Achterberg  {\it et al.} [The IceCube Collaboration],
  Phys.\ Rev.\  D {\bf 76} (2007) 027101
  [arXiv:0705.1781 [astro-ph]].

\bibitem[(Achterberg {\it et al.} 2007c)]{diffuse}
  A.~Achterberg  {\it et al.} [The IceCube Collaboration],
  Phys.\ Rev.\  D {\bf 76} (2007) 042008

\bibitem[(Achterberg {\it et al.} 2007c)]{grbpaper}
  A.~Achterberg  {\it et al.} [IceCube Collaboration and InterPlanetary Network],
   [arXiv:astro-ph/0705.1186], accepted for publication in Astrophysical Journal.

\bibitem[(Ackermann 2007)]{markus}
  M.~Ackermann {\it et al.} [IceCube Collaboration],
  {\it Prepared for 30th International Cosmic Ray Conference (ICRC 2007), M\'erida, Mexico, July 3 - 11, 2007}

\bibitem[(Aharonian 2007)]{Aharonian:2007bn}
  F.~Aharonian,
  arXiv:astro-ph/0702680.

\bibitem[(Braun 2007)]{jim_braun}
  J.~Braun {\it et al.} [IceCube Collaboration],
  {\it Prepared for 30th International Cosmic Ray Conference (ICRC 2007), M\'erida, Mexico, July 3 - 11, 2007}

\bibitem[(Cowen 2007)]{Cowen:2007ny}
  D.~F.~Cowen  [IceCube Collaboration],
  J.\ Phys.\ Conf.\ Ser.\  {\bf 60} (2007) 227.

\bibitem[(Finley 2007)]{chad}
  C.~Finley {\it et al.} [IceCube Collaboration],
  {\it Prepared for 30th International Cosmic Ray Conference (ICRC 2007), M\'erida, Mexico, July 3 - 11, 2007}

\bibitem[(Gerhardt 2007)]{lisa_gerhardt}
  L.~Gerhardt [IceCube Collaboration],
  {\it Prepared for 30th International Cosmic Ray Conference (ICRC 2007), M\'erida, Mexico, July 3 - 11, 2007}

\bibitem[(Gross 2007)]{big_andy}
  A.~Gross {\it et al.} [IceCube Collaboration],
  {\it Prepared for 30th International Cosmic Ray Conference (ICRC 2007), M\'erida, Mexico, July 3 - 11, 2007}

\bibitem[(Halzen 2002)]{Halzen:2002pg}
  F.~Halzen and D.~Hooper,
  Rept.\ Prog.\ Phys.\  {\bf 65} (2002) 1025
  [arXiv:astro-ph/0204527].

\bibitem[(Halzen 2007)]{Halzen:2007ip}
  F.~Halzen,
  Science {\bf 315} (2007) 66.

\bibitem[(Hill 2006)]{Hill:2006mk}
  G.~C.~Hill [IceCube Collaboration],
  arXiv:astro-ph/0611773.

\bibitem[(Hubert 2007)]{daan}
  D.~Hubert and A.~Davour [IceCube Collaboration],
  {\it Prepared for 30th International Cosmic Ray Conference (ICRC 2007), M\'erida, Mexico, July 3 - 11, 2007}

\bibitem[(Ishihara 2007)]{aya}
  A.~Ishihara [IceCube Collaboration],
  {\it Prepared for 30th International Cosmic Ray Conference (ICRC 2007), M\'erida, Mexico, July 3 - 11, 2007}

\bibitem[(Karle 2007)]{all_bracked}
  A.~Karle [IceCube Collaboration],
  {\it Prepared for 30th International Cosmic Ray Conference (ICRC 2007), M\'erida, Mexico, July 3 - 11, 2007}

\bibitem[(Resconi 2007)]{Resconi:2007nx}
  E.~Resconi [IceCube Collaboration],
  J.\ Phys.\ Conf.\ Ser.\  {\bf 60} (2007) 223.

\bibitem[(Song 2007)]{icetop}
  C.~Song {\it et al.} [IceCube Collaboration]
  {\it Prepared for 30th International Cosmic Ray Conference (ICRC 2007), M\'erida, Mexico, July 3 - 11, 2007}

\bibitem[(Tarasova 2007)]{oksana}
  O.~Tarasova [IceCube Collaboration],
  {\it Prepared for 30th International Cosmic Ray Conference (ICRC 2007), M\'erida, Mexico, July 3 - 11, 2007}

\bibitem[(Wikstrom 2007)]{gustav}
  G.~Wikstrom [IceCube Collaboration],
  {\it Prepared for 30th International Cosmic Ray Conference (ICRC 2007), M\'erida, Mexico, July 3 - 11, 2007}

\bibitem[(Wissing 2007)]{hickes}
  H.~Wissing [IceCube Collaboration],
  {\it Prepared for 30th International Cosmic Ray Conference (ICRC 2007), M\'erida, Mexico, July 3 - 11, 2007}

\end{thebibliography}
\end{document}